\documentclass[traditabstract]{aa}
\usepackage{amssymb,amsmath,amsfonts}
\usepackage{graphicx,graphics}
\usepackage[usenames]{color}

\newcommand{\beq}{\begin{equation}}
\newcommand{\eeq}{\end{equation}}
\newcommand{\bea}{\begin{eqnarray}}
\newcommand{\eea}{\end{eqnarray}}
\newcommand{\ks}{\kappa_{\rm S} }
\newcommand{\kt}{\kappa_{\rm T} }
\newcommand{\ras}{\mr{Ra_*} }

\newcommand{\nut}{{\mr{Nu_T}} }
\newcommand{\nutt}{Nu$_{\rm T}$ }
\newcommand{\nus}{{\mr{Nu_S}} }
\newcommand{\nust}{Nu$_{\rm S}$ }

\newcommand{\ehm}{ \color{black}}
\newcommand{\mr}{\mathrm}
\newcommand{\Nt}{  \tilde{\mathrm{N}} \mathrm{u} _\mathrm{T} }
\newcommand{\Nts}{ \tilde{\mathrm{N}} \mathrm{u}_\mathrm{S} }
\newcommand{\kse}{\kappa_\mr{S\, eff}}
\newcommand{\tbf}{}

\begin{document}
\title{Semiconvection: theory}
\titlerunning{Semiconvection}

\author{H.C.\ Spruit}

\institute{
  Max-Planck-Institut f\"{u}r Astrophysik,
  Karl-Schwarzschild-Str.\ 1,
  D-85748 Garching, Germany 
}
\date{\today}

\abstract{A model is developed for the transport of heat and solute in a system of double-diffusive layers under astrophysical conditions (viscosity and solute diffusivity low compared with the thermal diffusivity). The process of formation of the layers is not part of the model but, as observed in geophysical and laboratory settings, is assumed to be fast compared to the life time of the semiconvective zone. The thickness of the layers is \tbf{a} free parameter of the model. When the energy flux of the star is specified, the effective semiconvective diffusivities are only weakly dependent on this parameter. An estimate is given of the evolution of layer thickness with time in a semiconvective zone. The model predicts that the density ratio has a maximum for which a stationary layered state can exist, $R_\rho\la \mr{Le}^{-1/2}$. Comparison of the model predictions with a grid of numerical simulations is presented in a companion paper. 

\ehm
}

\maketitle

{\tt \obeylines }

\section{Introduction}

In stellar evolution, `semiconvection' denotes the situation where a thermally unstable stratification is stabilized against (adiabatic) overturning by a gradient in composition \tbf{(called {\it solute} in the following;} typically the Helium concentration, increasing with depth). It was first recognized as a complication in the calculation of stellar structure by R.~J.~Tayler (1953). Uncertainty whether the stabilizing gradient should be ignored (`according to Schwarzschild'),  included in the condition for overturning (`according to Ledoux'), or something in between has led to a number of different recipes for mixing of the Helium gradient. The evolution of the star subsequent to the semiconvective phase is sensitive to these differences (e.g. Weiss 1989, \tbf{Langer et al. 1989}, Langer 1991, Stothers \& Chin 1994, Langer and Maeder 1995). The fluid mechanics encountered in geophysics in the same case of a thermally unstable stratification stabilized by a stabilizing solute is called double diffusive or thermohaline convection. 

The observations show that such a stratification forms a stack of many thin layers, \tbf{called a `staircase',} each consisting of overturning convection sandwiched between stable steps in composition and temperature \tbf{(Turner \& Stommel 1964, Padman \& Dillon 1987, Schmid et al. 2010 and references therein)}. In effect, this is a `compromise between Schwarzschild and Ledoux'. Correspondingly, the net transport coefficients (of heat and solute) are intermediate between those of convection and  diffusion. 

The reason for this layer formation are understood (for references, cf. Spruit 1992,  hereafter S92, and Zaussinger \& Spruit 2013, hereafter ZS13). \tbf{The main theoretical contribution in this context has been the work of Proctor (1981), who showed analytically that a finite amplitude layered state exists for conditions when the stratification is still linearly stable: layering is the result of a {\it subcritical} instability. His analysis applies to the case of vanishing diffusivity of the solute, and Prandtl number not exceeding $\mathcal{O}(1)$, which is the astrophysically relevant case. In this limit, layered states exist whenever the Rayleigh number exceeds the critical value for normal convection, {\it independent of the strength of the stabilizing solute gradient}. An energy argument (S92) shows that the layered state in this case can be reached with an initial perturbation of vanishing amplitude as the layer thickness decreases.}

Attempts have been made to address the astrophysical problem with direct numerical simulations (Merryfield 1995, Biello 2001, Rosenblum et al. \tbf{2011}). This encounters two problems: the very high thermal diffusivity in a stellar interior (very small Prandtl number) cannot be matched without some form of approximation. More importantly  the quantity of main interest, the effective mixing rate, depends on the thickness of the double diffusive layers formed, a quantity that is not a stable outcome of the simulations. 

It makes sense to disentangle the `semiconvection problem' into two parts:  on the one hand the physics that determines the thickness of the layers,  on the other hand the effective mixing rate for a given layering state. This separation is especially meaningful because observations in geophysics and laboratory experiments show layer thickness to be a slowly changing function of time \tbf{compared with the overturning times within the layers}. In the following we concentrate on the second question, that is, the mixing rate is studied as a function layer thickness. 

It turns out that the simple observation of a layer structure consisting of an overturning zone between stable zones is sufficient to derive a predictive model for the effective transport coefficients (Sect.\  \ref{assum}). It is sufficiently quantitative to be tested against the results from numerical simulations. The translation of the model to the astrophysical case of a compressible stratification can be done exactly in the limit where the layers are thin compared with the pressure scale height. It is given in sec. \ref{astro}, and yields an easily implementable prescription for the mixing rate.

In the present treatment, the transition in composition and temperature between neighboring layers has a finite width, as opposed to S92, ZS10, where it was approximated as a step. This generalization turns out to make no significant difference for practical astrophysical application, but is essential for meaningful comparison of the theory with numerical results. Section \ref{evol} finally gives a (less quantitative) estimate of the layer thickness and its evolution in time.

The astrophysical term `semiconvection' will be used in the text interchangeably with the terms `double diffusive' or `thermohaline' convection used in laboratory and geophysical literature\footnote{Existing nomenclature is somewhat ambiguous. The opposite case of a destabilizing composition gradient in a stable thermal gradient is called `saltfingering', but  in geophysics also `thermohaline convection', where our semiconvective case is often called `diffusive convection' (e.g. Schmitt 1994). `Double diffusive' is usually meant to cover both cases.}.

\section{Physics of a semiconvective layer}
\subsection{Notation and definitions}
\label{notat}
As in the above, we will use the term `layer' for an individual double-diffusive step in the semiconvective \tbf{staircase}, and the term `solute' for the stabilizing component (e.g.\ Helium in Hydrogen). Such a layer consists of a zone of overturning convection between adjacent stagnant zones. In the stagnant zones overturning convection is suppressed by a stable ($N^2>0$) density gradient, and transport of heat and solute takes place by diffusion. 

\begin{figure}
\center\includegraphics[width=0.7\hsize]{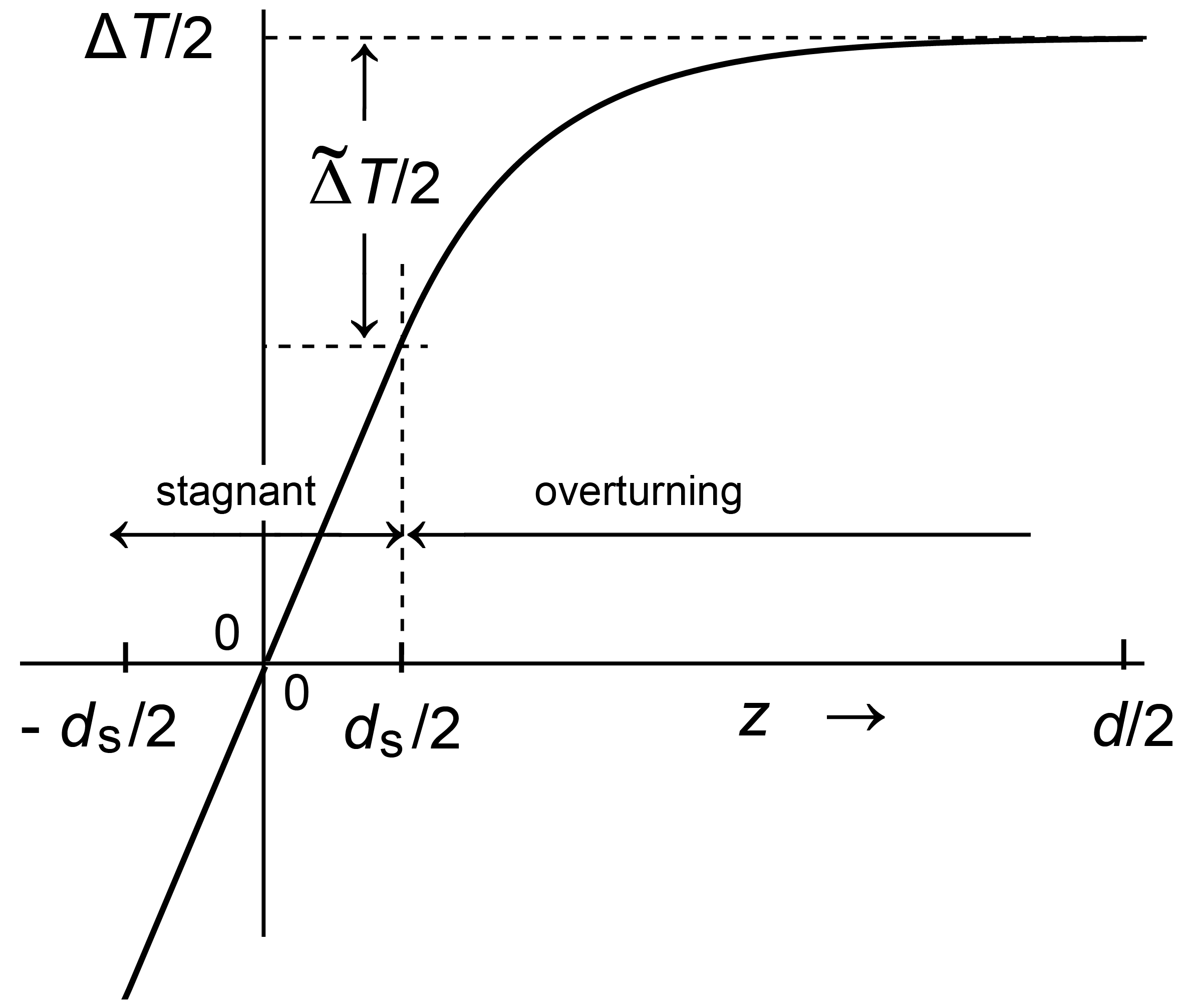}
\caption{Notation used, showing temperature as a function of depth through half of a semiconvective layer of thickness $d$,  from the middle ($z=0$) of the stagnant zone (of thickness $d_s$), to the middle of the overturning zone at $z=d/2$.  $\Delta T$ is the temperature difference across the whole layer, $\tilde\Delta T$ across the overturning zone. The solute profile has the same shape but different amplitudes $\Delta S$ and $\tilde\Delta S$.}\label{sketch}
\end{figure}

As illustrated in Fig.\ \ref{sketch}, depth through the layer is counted from the middle of a stagnant zone. The thickness of the layer as a whole is $d$, that of the stagnant zone $d_{\rm s}$. The temperature difference across the layer is $\Delta T$, the solute difference $\Delta S$. 
The stabilizing influence of the solute is expressed conveniently in terms of the density ratio, the ratio of density differences caused by temperature and solute differences $\Delta T$ and $\Delta S$ across the layer:
\beq R_\rho=\beta\Delta S/\alpha\Delta T, \label{rrho}\eeq
where $\alpha$,\tbf{ the thermal expansion coefficient, is the relative density decrease per unit temperature and $\beta$, called the {\it haline contraction coefficient}, is  the relative density increase per unit of solute concentration}. Since density increases with $S$ and decreases with $T$, the density differences are of the opposite sign when $\Delta T$ and $\Delta S$ both increase with depth (in the direction of gravity)  \tbf{as in our semiconvective case}. 

The flow in the overturning zone of the  layer is driven by a temperature difference $\tilde\Delta T$ ($<\Delta T$) and opposed by a solute difference $\tilde\Delta S$  ($<\Delta S$). Associated with these is an `internal density ratio', related to the overall density ratio by
\beq \tilde R_\rho\equiv\beta\tilde\Delta S/\alpha\tilde\Delta T= R_\rho {\tilde\Delta S\over \Delta S}{\Delta T\over \tilde\Delta T}.\eeq 
Obviously, the values of $\tilde\Delta T$ and $\tilde\Delta S$ depend on where we define the boundaries between stagnant zone and overturning zone. Since the internal layer is actually convecting, the boundary has to be set at a point in the $T,S$ profile where $\tilde R_\rho <1$ (see also Sect.\ \ref{assum}). 

The overturning zone is characterized by a Rayleigh number; for convection in a layer of thickness $D$, with temperatures $T_{\rm t}$ and $T_{\rm b}$ at top and bottom, \tbf{the Rayleigh number} is
\beq \mr{Ra}={g\alpha(T_{\rm b}-T_{\rm t})D^3\over\kt\nu},\label{Ra}\eeq
where $g$ is the acceleration of gravity, $\kt$ the thermal diffusivity, and $\nu$ the (kinematic) viscosity. In our case (Fig.\ \ref{sketch}), $D$ has the value $d-d_\mr{s}$. The critical value for onset of convection \tbf{is} of order Ra$_{\rm c}=1400$ (for no-slip boundary conditions). If viscosity is low, (Prandtl number Pr$=\nu/\kt\ll 1$), the heat flux at large Rayleigh number becomes {\em independent} of viscosity, (a fact that is used implicitly in the `mixing length' formalism for convection in a stellar interior). At high Ra, a  more relevant quantity to characterize the heat flux in this case is the modified Rayleigh number Ra$_*$:
\beq \ras=\mr{Pr\,Ra}={g\alpha(T_{\rm b}-T_{\rm t})D^3\over\kt^2}.\label{Ra*}\eeq
Its square root can be read as the ratio of the thermal diffusion time $D^2/\kt$ to the free fall time of the density contrast $\alpha(T_{\rm b}-T_{\rm t})$  over the distance $D$.

Let $F$ be the (time \tbf{averaged}) heat flux across the layer, and $F_\mr{d}$ the flux in the absence of convection, i.e. when the temperature profile between $T_\mr{b}$ and $T_\mr{t}$ is determined by diffusion only. The Nusselt number  is then defined as
\beq \nut=F/F_\mr{d}. \eeq
Similarly, there is a Nusselt number for the solute flux $F_\mr{S}$:
\beq \nus=F_\mr{S}/F_\mr{Sd}. \eeq
In the absence of a solute, i.e. for normal (unstratified) laboratory convection, \nutt is a function of Pr and Ra only (apart from boundary conditions and geometry). In the case of semiconvection,  the Nusselt numbers are functions of two additional parameters characterizing the solute:  the density ratio $R_\rho$ and the Lewis number Le, the ratio of solute diffusivity $\ks$ and thermal diffusivity $\kt$:
\beq \mr{Le}= \ks/\kt.\eeq

\subsection{Transport coefficients: classical results}
The heat flux is determined by the boundary layers at top and bottom. As the overturning flow passes along the boundary, the \tbf{temperature contrast with the} boundary diffuses into the flow, and it is this temperature difference that carries the heat flux. If  $\tau$ is the time during which the flow is in contact with the boundary before descending/ascending into the bulk, the depth $D_\mr{T}$ over which the temperature difference penetrates is $D_\mr{T}= (\kt\tau)^{1/2}$. At high Rayleigh numbers, this depth is small compared with the layer thickness $d$.  In a simple two-dimensional view of the flow the amount of fluid carrying this difference also scales with $D_\mr{T}$. The same flow along the boundary determines how much solute contrast diffuses into the flow. The solute diffusion depth is thus  $D_\mr{S}= (\ks\tau)^{1/2}$, and \tbf{the} amount of solute flowing into the overturning zone is proportional to $D_\mr{S}$ \tbf{and inversely proportional} to the flow speed $v$ along the boundary. 
 
The ratio of the convective fluxes of solute and heat is thus expected to scale as $(\ks/\kt)^{1/2}$.  On the other hand, the (diffusive) fluxes over the thickness of the overturning layer,  in the absence of convection, scale as $\ks$ and $\kt$. In terms of the Nusselt numbers, this implies, in the limit  $\nut,\,\nus\gg 1$ such that the boundary layers are thin compared with the layer thickness: 
\beq \nus=q\,\mr{Le}^{-1/2} \nut,\label{class}\eeq 
where $q$ is a numerical factor expected to be of order unity.
If the heat flux is kept fixed, the effective diffusivity of the solute then scales as 
\beq \kappa_\mr{s, eff}=\ks\nus=q\, (\ks\kt)^{1/2}\nut.\eeq
These are classical scalings, as found in previous analyses of double diffusive convection, e.g. in Turner (1980, 1985). In this form they do not address the dependence of the fluxes on the density ratio R$_\rho$. The argument above implicitly assumes that all the density contrast in the solute boundary layer $D_\mr{S}$ is carried across with the flow. In reality, only a fraction of the boundary layer can flow across, namely the fraction that is not  too buoyant to be carried down by the flow, respectively too heavy to be carried up. Taking this into account (Spruit 1992, hereafter S92), yields a correction to (\ref{class}):
\beq \nus={q\over R_\rho} \mr{Le}^{-1/2}\nut.\qquad (\nut\gg 1) \label{s92}\eeq 
This scaling applies to the overturning zone; it would be valid for the layer as a whole only if the stagnant zone were absent. More precisely, it assumes that the thickness $d_\mr{S}$ of the stagnant zone is smaller than the thickness of the boundary layers $D_\mr{T},D_\mr{S}$. This is not always the case (cf. Sect.\  \ref{fixed} below). In the following, the presence of the stagnant zone is taken into account explicitly, by \tbf{consistently taking} the distinction between between Nusselt numbers for the overturning zone and those for the layer as a whole \tbf{into account}. This results in a generalization of (\ref{s92}).

\subsubsection{`Erosion'}

To determine the numerical factor $q$ in (\ref{s92}) more quantitatively than order unity, the hydrodynamic interaction of the flow with the stagnant zone has to be considered in more detail. This is somewhat beyond the scope of the model to be developed here, but we can identify a process involved, and use this to show that $q$ is probably somewhat larger than unity.  I will call $q$ the `erosion factor'. 

The unsteady flow in the overturning zone induces perturbations in the stagnant zone. Since it is stably stratified, these take the form of internal gravity waves. These can contain internal structure on length scales (perpendicular to the interface) less than the thickness of the stagnant zone. Diffusion of solute on these length scales increases the amount that has sufficiently low buoyancy to be carried with the overturning flow, hence we may expect $q>1$. In the absence of a more detailed theory, its value can in principle be used as a fitting parameter, as long as it is not taken larger than order unity. In the following, however, I will ignore this option, and simply set 
\beq q=1 .\eeq 
For astrophysical applications, the effective solute transport will turn out to be so low that tuning the erosion factor by order unity would have little effect anyway. 
At low Le, this erosion process takes place well inside the thermal boundary layer. It consequently affects only the transport of the solute; its effect on the transport of heat can be neglected.

\section{Model}

\label{assum}
The layer of thickness $d$ now consists (Fig.\ \ref{sketch}) of a stagnant zone of thickness $d_\mr{S}$, and a overturning zone occupying the remainder of $d$. An incompressible (Boussinesq) fluid is assumed, to be generalized to the compressible astrophysical case in Sect.\  \ref{astro}. 

In the stagnant zone, we assume \tbf{that} the transport of heat and solute is by diffusion only (i.e. ignoring the possible `erosion' effect discussed above). In the overturning zone of the layer, the flow is approximated as convection \tbf{as it would take place} in the absence of a solute. The presence of the solute affects the flow somewhat, but the amount of solute carried, and hence its influence on the flow, vanish in the limit of low solute diffusivity (S92, Schmitt 1994). Apart from the boundary conditions it sets, the stagnant zone has little effect on the flow inside the overturning zone. In the limit $\mr{Le}\ll 1$ considered here, the effect of the solute on the flow of heat in the overturning zone  can thus be neglected.

To describe transport of heat in the overturning zone, we use a fit from laboratory measurements for the Nusselt number  as a function of the temperature difference. Since the thickness and temperature difference are different from those of the layer as a whole, the Rayleigh and Nusselt numbers of the overturning zone are distinguished here with a $~\tilde{ }$\ . With the notation
\beq \epsilon=\tilde\Delta T/\Delta T,\quad \delta= d_{\rm{s}}/d,\eeq
they are related to Ra and \nutt by (cf. eq. \ref{Ra*} and the geometry sketched in Fig. \ref{sketch}):
\beq \tilde{\mr {R}}\mr{a}_*=\ras\,\epsilon(1-\delta)^3,\label{rasra}\eeq
\beq \Nt= \nut~(1-\delta)/ \epsilon.\label{ntn}\eeq

In the measurements of Niemela et al. (2000), at Ra up to $10^{17}$, the Nusselt number is well fit by a power law of slope 0.309 and amplitude 0.124. The Prandtl number in these experiments is order unity, so Ra$_*$ and Ra are approximately the same. 
Combining the above, I approximate the heat flux in the overturning zone as given by
\beq \Nt=1+a(\tilde{\mr {R}}\mr{a}_*-\mr{Ra}_{*c})^b, \quad (a=0.124,~b=0.309).\label{niem}\eeq
Since, as argued in \ref{notat} the dependence on Prandtl number is expected to be weak below Pr$=1$, (\ref{niem})  will be assumed approximately valid for all Pr $\le 1$ (this can be checked with numerical simulations, see \tbf{Zaussinger and Spruit 2013, hereafter ZS13}). The constant 1 and the critical Rayleigh number have been added to better approximate the behavior at low Ra$_*$.

The buoyancy of the solute is opposite to that of the driving temperature difference, and the flow will only be driven by density differences of the unstable sign. This implies that the solute concentration contrast in the overturning flow, $\tilde S$ say, is limited by the temperature contrast $\tilde T$ through 
\beq R_\rho\tilde S/\Delta S\le \tilde T/\Delta T.\label{rel}\eeq
At the boundary between the overturning and the diffusive zones, these contrasts are $\tilde S=\tilde\Delta S$, $\tilde T=\tilde\Delta T$. Define then the location of this boundary as the point where the equals sign holds in (\ref{rel}), that is, the point where the buoyancy of the solute is just low enough for it to be carried with the thermally driven flow in the overturning zone (cf. discussion above). This yields the relation
 \beq {\tilde\Delta S\over\Delta S}= {1\over R_\rho}{\tilde\Delta T\over\Delta T}.\label{buo}\eeq
In the overturning  zone, the original scaling (\ref{class}) applies, i.e.  $\Nts= \mr{Le}^{-1/2}\Nt$ (assuming $q=1$), since $\tilde\Delta S$ is the density difference that can just be carried with the flow. 
At low Ra, $\Nts$ should approach unity at the same time as $ \Nt$. To accomplish this I adopt a slight modification:
\beq  \Nts-1= \mr{Le}^{-1/2}(\Nt-1),\label{nusnut}\eeq
which is now assumed to hold for all $\Nt>1$. 

This completes the definition of the model. It defines the fluxes of heat and solute uniquely as functions of the external parameters.

\section{Effective diffusivities}
\label{effd}

\begin{figure}
\center\includegraphics[width=0.7\hsize]{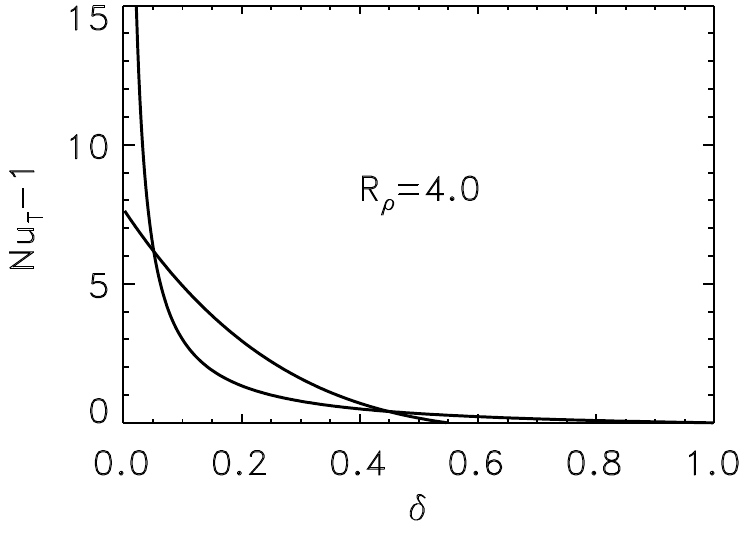}\\
\center\includegraphics[width=0.7\hsize]{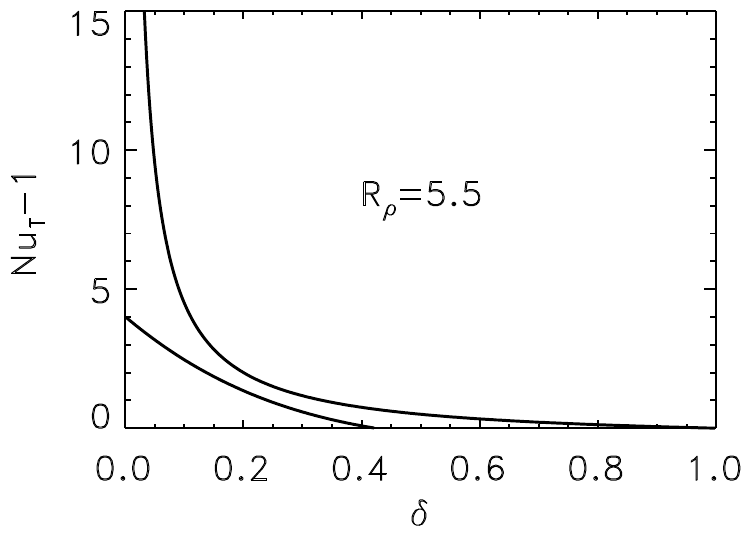}
\caption{Determination of the thickness $\delta$ of the stagnant zone, (example for Ra$_*=10^7$, Le$=0.01$, Pr$=0.1$).  The steep curve shows the heat flux in the stagnant zone as a function of the assumed value of $\delta$, the shallow curves the corresponding flux in the overturning zone. Intersection points are possible values of $\delta$ for stationary heat flow. Top: density ratio $R_\rho=4$. For the higher density ratio of 5.5 (bottom), there is no value of the thickness for which the two match (see text).}\label{dsdep}
\end{figure}

In a stationary state,  the fluxes of heat and solute are constant with depth through the layer. The fluxes by diffusion in the stagnant zone are equal to the fluxes in the overturning zone. Expressing this in terms of the Nusselt numbers, we first need to write $\Nt$ and $\Nts$ in terms of the Nusselt numbers for the layer as a whole. For temperature this is given by (\ref{ntn}):
\beq \nut=\Nt{\epsilon\over(1-\delta)}.\label{nd}\eeq
The equivalent relation for the solute is
\beq \nus=\Nts{\epsilon_\mr{S}\over(1-\delta)},\label{nsd}\eeq
where (using \ref{buo}):
\beq \epsilon_\mr{S}\equiv\tilde\Delta S/\Delta S=\epsilon/R_\rho. \eeq
The fluxes in the stagnant zone are given by:
\beq \nut=(1-\epsilon)/\delta, \qquad\nus=(1-\epsilon_\mr{S})/\delta.\label{nudiff}\eeq
(since carried by diffusion alone). With (\ref{nusnut}), eqs. (\ref{nd})-(\ref{nudiff}) yield two equations for the unknown $\delta$ and $\epsilon$:
\beq (1/\delta-1)(1/\epsilon-1)=\Nt,\label{nnn}\eeq
\beq (1/\delta-1)(R_\rho/\epsilon-1)=1+\mr{Le}^{-1/2}(\Nt-1),\label{nns}\eeq
with $\Nt$ given by (\ref{niem}).
Eliminating $\Nt$ between these two yields
\beq 1-\delta-\epsilon/R_\rho=(1-\delta-\epsilon)/Q,\label{eps}\eeq
where
\beq Q=R_\rho\mr{Le}^{1/2}.\label{Q}\eeq
Solving this for $\epsilon$:
\beq \epsilon=(1-\delta){1-Q\over 1-Q/R_\rho}.\label{eps1}\eeq
Since $\epsilon$ must be a positive number, and $R_\rho$ is larger than 1 (otherwise we would not be `below Ledoux'), it follows that in a stationary state as envisaged $Q$ must be less than 1, or $R_\rho<\mr{Le^{-1/2}}$. This is a necessary condition (independent of the Rayleigh number), but not sufficient. The actual, somewhat lower, value of the critical density ratio has to be determined from (\ref{nd})-(\ref{nsd}) by solving for $\delta$ as well as $\epsilon$. To see how this solution comes about, consider the left-hand and right-hand sides of (\ref{nnn}) separately. They represent the heat flux in the stagnant and the overturning zones, respectively, and in a stationary state \tbf{they} are equal.  Fig. \ref{dsdep} shows the two as functions of $\delta$, with $\epsilon$ taken from (\ref{eps1}) and $\Nt$ from (\ref{niem}). 

The two intersection points of the curves are potential solutions for a steady state. To see which of the two is the relevant one, consider the slopes of the curves. At the equilibrium point with the smaller value of $\delta$ the flux in the stagnant zone decreases more rapidly with $\delta$ than the flux in the overturning zone. A small decrease of $\delta$ away from the equilibrium would increase the flux in the stagnant zone relative to that in the overturning zone. This would result in a temperature deficit at the boundary between the two, causing  the temperature to decrease there. This would reduce the heat flux in the stagnant zone again, the opposite of the assumed perturbation. This equilibrium point is thus the stable one; the intersect\  at the larger $\delta$ is an unstable point. 

\tbf{The necessary condition for existence of the layered state, $R_\rho<\mr{Le}^{-1/2}$ also figures prominently in Proctor (1981). In his analysis it shows up as a necessary condition for its validity.}

\subsection{Maximum density ratio}
\label{rrhmax}
\begin{figure}
\center\includegraphics[width=0.8\hsize]{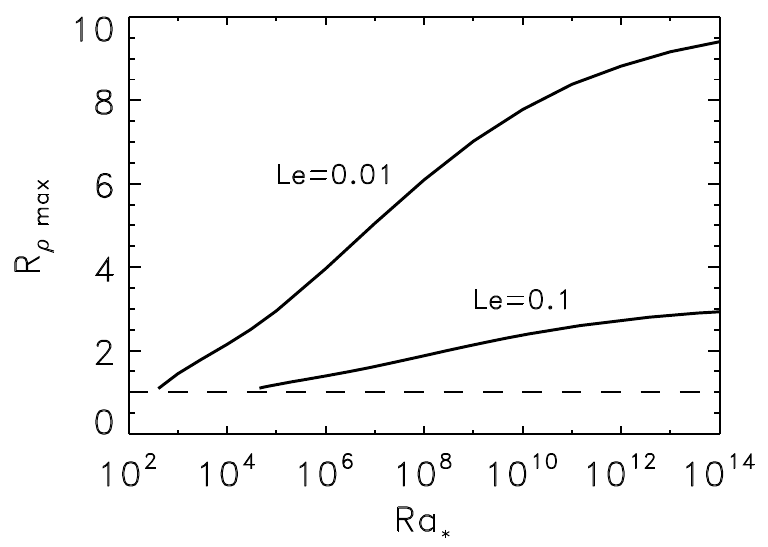}\\
\caption{Maximum value of the density ratio, as a function of the modified Rayleigh number of the layer, for $\mr{Le}=0.01$ and $\mr{Le}=0.1$.}\label{rrhom}
\end{figure}

The maximum density ratio can be determined as a function of Ra$_*$ as the value for which the two curves in Fig.\ \ref{dsdep} just touch. The result is shown in Fig. \ref{rrhom}. At large Ra$_*$, $R_{\rho\,\mr{max}}$ (slowly) approaches  the value $\mr{Le}^{-1/2}$. The behavior of \nutt and $\delta$ near $R_{\rho\,\mr{max}}$ is shown in Fig. \ref{rrhdep} for an illustrative case.

At the critical density ratio the value of $\delta$ is about 0.2  (cf.\ Fig. \ref{dsdep}), while the Nusselt number reaches a minimum value of the order of a few. At this density ratio, the model predicts a jump from an overturning state ($\nut >1$) to a purely diffusive state $\nut=1$. An example is shown in Fig.\ \ref{rrhdep}. \tbf{The presence of this jump suggests that the behavior of the system near the maximum density ratio needs a closer look than is possible with the present model. This question is explored with numerical simulations in ZS13.}

The maximum on $R_\rho$ can also be read as a minimum on the Rayleigh number. For an observed density ratio of 4 at Le$=0.01$ for example, one should not expect to see double-diffusive layering with layer thickness less than corresponds to a Ra$_*$ of $10^6$ (\tbf{from} Fig.\ \ref{rrhom}). In terms of the reference length $d_0$ defined below (Eq. \ref{d0}), the layers must have \tbf{a thickness} $d/d_0>45$, for this combination of $R_\rho$ and Le.

\subsection{Solute flux}
\label{solu}
\begin{figure}
\center\includegraphics[width=0.8\hsize]{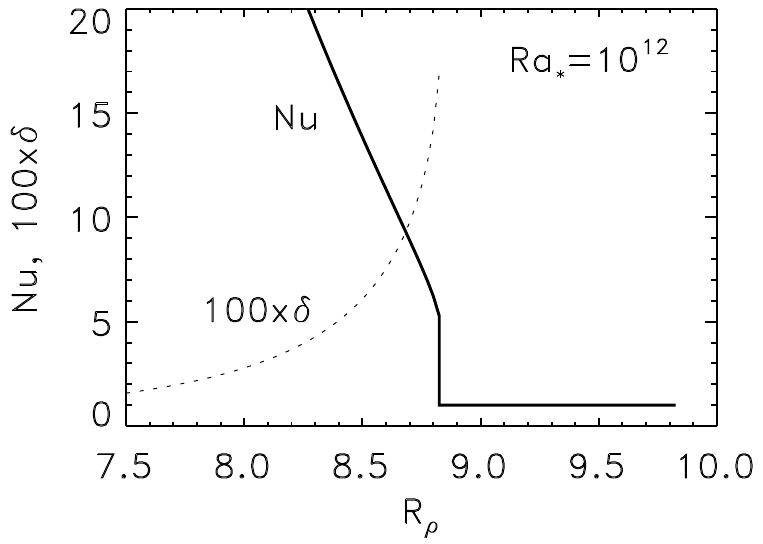}\\
\caption{Dependence of Nusselt number and thickness of the stagnant zone on density ratio R$_\rho$ in a range around the  critical value,  for $\mr{Le}=0.01$, Ra$_*=10^{12}$ (R$_{\rho\, \mr{max}}=8.82$).\label{rrhdep}}
\end{figure}

The net solute Nusselt number can be expressed in terms of \nutt. Using (\ref{nusnut})--(\ref{nudiff}) and (\ref{Q})--(\ref{eps1}), this yields the simple expression
\beq \nus=1+{1\over R_\rho \mr{Le}^{1/2}}(\nut-1).\label{nunus}\eeq
The stagnant zone of finite thickness included here thus leads to the same relation between \nust and \nutt as the simpler model in S92 (cf. eq. \ref{s92} above). There is a difference, however, in the relation between \nutt and Ra$_*$, and in the presence of a maximum density ratio (cf.\ Figs.\ \ref{rrhom},\ref{rrhdep}).

\subsection{Asymptotic behavior}
 The Nusselt number is, with (\ref{nd}), (\ref{eps1}):
\beq \nut=\Nt\, P\eeq  
where
\beq P\equiv {1-Q\over 1-Q/R_\rho}={1-\mr{Le}^{1/2}R_\rho\over1-\mr{Le}^{1/2}},\eeq
and $\Nt$ is given in terms of $\tilde{\mr {R}}\mr{a}_*$ by (\ref{niem}). For $\mr{Ra}_*\gg 1$, and with $R_\rho$ not close to $R_{\rho\, \mr{max}}$, the stagnant zone is thin, $\delta\ll 1$. $\tilde{\mr {R}}\mr{a}_*$ as a function of $R_\rho$ and Le is then, with (\ref{ntn}), (\ref{eps1}): 
\beq \tilde{\mr {R}}\mr{a}_*\approx \mr{Ra_*}\, P.\qquad (\mr{Ra_*}\gg 1, ~R_\rho<R_{\rho\,\mr{max}})\eeq
The solute Nusselt number follows from (\ref{nunus}),
\beq \nus\approx{\nut\over R_\rho \mr{Le}^{1/2}}.\qquad (\mr{Ra_*}\gg 1, ~R_\rho<R_{\rho\,\mr{max}})\label{nunusa}\eeq

It is instructive to compare $\delta$ to the thickness of the boundary layers $D_\mr{T}$, $D_\mr{S}$ of the overturning zone. These can be written in terms of the Nusselt numbers \nutt and \nust since at the boundaries with the stagnant zone the fluxes are carried by diffusion. For the thermal boundary layer for example this implies
\beq D_\mr{T}/(d-d_\mr{S})=1/\Nt. \eeq
Using (\ref{nnn}), (\ref{eps1}):
\beq \Nt=(R_\rho/P-1)/\delta,\eeq
hence in the limit $\delta\ll 1$:
\beq \delta\approx {D_\mr{T}\over d}({R_\rho\over P}-1)={D_\mr{T}\over d}{R_\rho-1\over 1-Q}.\label{blt}\eeq
Unless $Q$ is close to unity, the width of the stagnant zone is thus of the same order as the boundary layers of the overturning zone, and the distinction between the two becomes somewhat academic. At large density ratio, or under the fixed heat flux conditions discussed in the next Section, however, the distinction becomes more significant. 

\section{Fixed heat flux conditions}
\label{fixed}
In laboratory situations and theoretical analyses it is usual to consider the temperature difference or temperature gradient as given and the heat flux or Nusselt number as the object to be determined. In some natural systems, however, the conditions under which double diffusive convection occurs are closer to ones where the heat flux is the given quantity. In the east African volcanic lakes, for example, the heat flux imposed by the influx at the bottom of the lake is probably more of a given than the temperature at the bottom of the lake. The same is the case in semiconvective zones of stars. 

The temperature difference across the layers adjusts to an imposed heat flux. Since both the density ratio and the Rayleigh number depend on $\Delta T$, neither of these can be used as control parameter any more. This requires a change of perspective on the problem. 

The stratifications of both temperature and solute change with time, under the effective transport properties of the semiconvective process. In the limit of low solute diffusivity, the time scale for changes in the solute profile is long compared with the time scale on which the temperature profile adjusts. In this \tbf{thermally} quasisteady state, the solute profile can be taken as fixed.  Assume therefore that the mean solute gradient $\mr{d}S/\mr{d}z$ is given. As new control parameters, use the heat heat flux $F$ and layer thickness $d$.

\begin{figure*}
\center\includegraphics[width=0.7\textwidth]{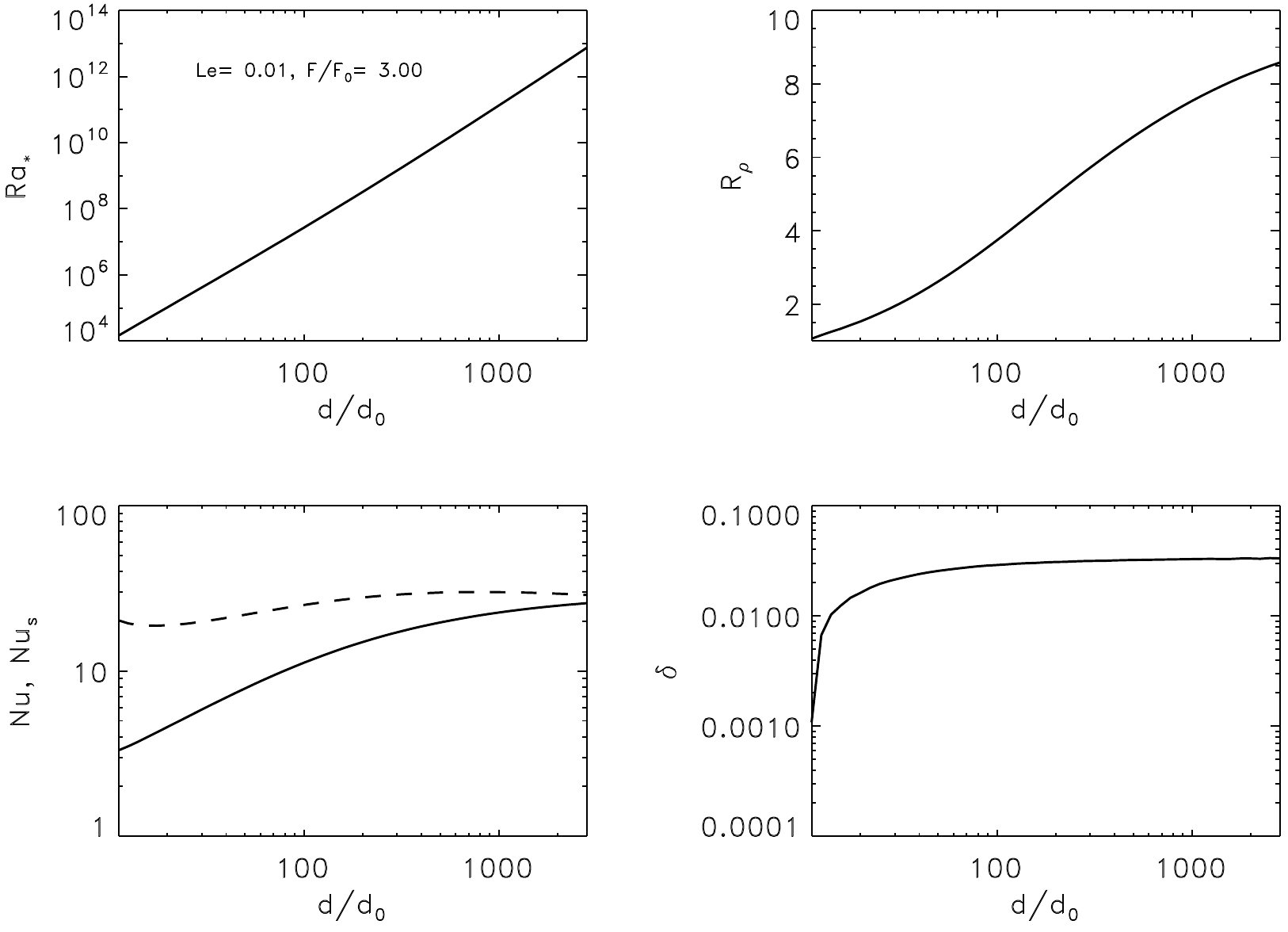}\\
\caption{Dependences on layer thickness $d$, for fixed heat flux conditions ($F/F_0=1.5$). In reading order: Rayleigh number, density ratio, Nusselt numbers (dashed for solute) and thickness of the stagnant zone.
\label{fixf}}
\end{figure*}

To make the imposed heat flux practical as control parameter, measure it with respect to a reference flux $F_0$ that can be expressed in terms of the solute gradient. Take for this the diffusive heat flux  that would be present in the linear temperature profile $T_0(z)$ that is just marginally stable against adiabatic overturning (`Ledoux'). If $K$ is the thermal conductivity, the heat flux of this stratification is
\beq  F_0=K{\mr{d}T_0\over\mr{d}z},\eeq
and its density ratio is
\beq R_{\rho\,0}=\beta{\mr{d}S\over\mr{d}z}/\alpha{\mr{d}T_0\over\mr{d}z}.\eeq
Since we have assumed that the stratification is marginally stable, $R_{\rho\,0}=1$. The reference heat flux is thus
\beq F_0=K{\beta\over\alpha}{\mr{d}S\over\mr{d}z}.\eeq
 The actual layered state has a Nusselt number \nutt and a density ratio $R_\rho>1$; by definition of the Nusselt number, its heat flux is
\beq F=\nut\,K{\Delta T\over d}. \eeq
With $\Delta S=d\,\mr{d}S/\mr{d}z$:
\beq F/F_0=\nut{\alpha\Delta T\over\beta\Delta S}=\nut/R_\rho. \label{ff0}\eeq
To replace the Rayleigh number as control parameter, note that it can be written in terms of solute gradient and density ratio as (using \ref{rrho}) 
\beq \mr{Ra_*}={g\beta\over\kt^2R_\rho}d^4{\mr{d}S \over\mr{d} z}. \eeq
The quantity
\beq d_0=({\kt^2\over g \beta}/{\mr{d}S\over\mr{d}z})^{1/4}=(\kt/N_{\rm S})^{1/2}, \label{d0}\eeq
is a characteristic length scale of the problem: the distance \tbf{over which temperature diffuses on the buoyancy time scale} of the stable solute gradient, $N_{\rm S}^{-1}=(g\beta\,\mr{d}S/\mr{d}z)^{-1/2}$. The Rayleigh number can then be written as
\beq \mr{Ra_*}={1\over R_\rho}(d/d_0)^4.\label{rad}\eeq
The Nusselt number in (\ref{ff0}) is a function of $R_\rho$ and Ra$_*$, which can be evaluated from Eqs.\ (\ref{niem},\ref{nudiff},\ref{nnn},\ref{nns}) in Sect.\  \ref{effd} above. Together, eqs. (\ref{ff0}) and (\ref{rad}) thus determine Ra$_*$ and R$_\rho$, and other quantities of interest, as functions of the new control parameters $d/d_0$ and $F/F_0$. An example of the dependence on $d/d_0$ of Ra$_*$, $R_\rho$, the Nu's, and $\delta$ is shown in Fig.\ \ref{fixf} for $F/F_0=3$.

\subsection{Asymptotic dependences for fixed heat flux}

The limiting case $d/d_0\gg 1$ (corresponding to Ra$_*\gg 1$), has a pleasingly simple form. As expected from the discussion in Sect. \ref{effd} (cf.\ Fig.\ \ref{rrhom}), the density ratio approaches its maximum value Le$^{-1/2}$ in this limit. With (\ref{ff0}) and (\ref{nunus}) the Nusselt numbers approach the value
\beq \nus\approx \nut\approx \mr{Le}^{-1/2}F/F_0 \qquad (d/d_0\gg 1). \label{asy}\eeq
 For $\nus$, this is in fact a good approximation also at lower values of $d/d_0$, as Fig.\ \ref{fixf} (lower left) shows. The corresponding effective solute diffusivity becomes
\beq \kappa_\mr{s,\, eff}=\nus\ks\approx (\ks\kt)^{1/2}\,F/F_0,\label{kseff}\eeq
in agreement with the classical `geometric mean of diffusivities' scaling.
The thickness of the stagnant zone is related to $\nus$ by eq.\  (\ref{nudiff}). In the present limit, $\epsilon_\mr{S}\ll 1$, hence
\beq \delta\approx \nus^{-1} \approx \mr{Le}^{1/2}(F/F_0)^{-1}\quad (d/d_0\gg 1).\eeq
The relative thickness of the stagnant zone is thus small for low Lewis numbers or at high heat flux. In the  opposite case of modest Le {\em and} conditions closer to marginal, it can be \tbf{a} significant fraction of the layer thickness, however. This may be the explanation for the relatively large thickness of the stagnant zones observed in lake Kivu (Schmid et al. 2010). The heat flux measured there corresponds to  Nusselt numbers of order 2. \tbf{A stagnant zone thickness} $\delta$ of order 30\% is observed, larger than in other natural cases like the double-diffusive steps under the arctic ice sheet (e.g.\ Timmermans et al. 2008 and references therein). Though the present analysis does not apply directly to this case because the assumption Pr $<1$ does not hold (Pr $\approx 7$ for water), the thickness of the stagnant zones in  lake Kivu may be an indication that it is actually close to the marginal state to be expected at imposed low heat flux conditions. 

\subsubsection{Thickness of the stagnant zone}
\label{stag}
As Fig.\ \ref{fixf} shows, the density ratio approaches its maximum in the limit $d/d_0\gg 1$ (cf Sect. \ref{rrhmax}):
\beq R_\rho\approx \mr{Le}^{-1/2}\quad \rightarrow\quad 1-Q\ll1. \eeq
The actual value of $1-Q$ as a function of  the parameters $d/d_0$ and $F/F_0$ is a  somewhat complicated expression involving the coefficients $a,b$ in (\ref{niem}) \tbf{(it will not be needed in the following)}. 

Comparing $\delta$ with the thermal boundary layer thickness $D_\mr{T}$ of the overturning zone then yields, from (\ref{blt}) :
\beq  \delta\approx {D_\mr{T}\over d} {\mr{Le}^{-1/2}-1\over 1-Q}.\eeq
In this asymptotic case with fixed heat flux, the stagnant zone is thus much wider than the boundary layers of the overturning zone (but still thin compared with the layer thickness $d$). Its presence manifests itself \tbf{in} the approximate equality of \nutt and \nust (eq. \ref{asy}) as opposed to the original estimate (\ref{class}).

\section{Astrophysical conditions}
\label{astro}
For the compressible gas in a stellar interior, an incompressible approximation (Boussinesq) can still be used for flows on length scales small compared with the pressure scale height and time scales short compared with the sound travel time over this length, provided two factors are taken into account (e.g.\ Massaguer \& Zahn 1980). First, the adiabatic lapse rate $(\mr{d}T/\mr{d}z) _\mr{a}$ has to be subtracted from the temperature difference driving the flows. The modified Rayleigh number for a layer of thickness $d$ is now
\beq \mr{Ra_*}={g\alpha\over\kt^2}d^4({\mr{d}T\over\mr{d}z}-{\mr{d}T\over\mr{d}z}\vert_\mr{a}),\eeq
where $\alpha=1/T$ for the ideal gas with constant ratio of specific heats assumed here. In terms of the logarithmic gradient $\nabla=\mr{d}\ln T/\mr{d}\ln p$:
\beq \mr{Ra_*}={g d^4\over\kt^2H}(\nabla-\nabla_\mr{a}),\label{raa}\eeq
where $H$ is the  scale height of the gas pressure $p$,
\beq H=\mr{d}z/\mr{d}\ln p.\eeq 
This length scale is not present in the Boussinesq case. An equivalent of the quantity $d_0$ used in Sect.\ \ref{fixed} still plays a role as well, but in the astrophysical context $H$ is a more conventional choice as reference length. $\mr{Ra_*}$ can then be written as
\beq \mr{Ra_*}=f^2({d\over H})^4(\nabla-\nabla_\mr{a}),\eeq
where the dimensionless number $f$,
\beq f=(g H^3/\kt^2)^{1/2}, \eeq
typically a large number, is the ratio of the thermal diffusion and free fall time scales over a pressure scale height. The dimensionless layer thickness $d/H$ will be the control parameter replacing $d/d_0$ above.

A second difference concerns the radiative heat flux, since it is driven by the temperature gradient itself rather than the potential temperature\footnote{Note that in the presence of a solute the superadiabaticity $\nabla-\nabla_\mr{a}$ is not equivalent to the entropy gradient any more.  Instead, it measures the gradient of {\em potential temperature}, the temperature on adiabatic displacement, in pressure equilibrium, relative to a given reference level.}. This affects the definition of Nusselt number, as well as the choice of control parameter specifying \tbf{the} heat flux.
Define a thermal conduction constant  $k$ (a function of the opacity and the thermodynamic state of the gas), in terms of the the radiative  heat flux $F_\mr{r}$:
\beq F_\mr{r}=k\nabla . \eeq
Following standard notation, represent the total heat flux $F$ by the radiative gradient $\nabla_\mr{r}$:
\beq F=k\nabla_\mr{r}.\label{rgrad}\eeq
$\nabla_\mr{r}$ will be the \tbf{astrophysical} control parameter specifying heat flux.
The ratio of density changes due to composition and temperature changes upon adiabatic displacement in the stratification, i.e. the density ratio, is
\beq R_\rho={\nabla_\mu\over\nabla-\nabla_\mr{a}},\label{rrha}\eeq
where 
\beq \nabla_\mu=\mr{d}\ln\mu/\mr{d} \ln p,\eeq
and the $\mu$ the mean atomic weight per particle.
Define the Nusselt number \nutt through
\beq F=k\nabla_\mr{a}+\nut~k(\nabla-\nabla_\mr{a}), \label{fnu}\eeq
so that it measures the part of the heat flux that is due to (only) the superadiabatic part of the temperature gradient, instead of the total heat flux. This has to be clearly distinguished from the ratio $\nabla_\mr{r}/\nabla$ of heat flux to radiative heat flux in the stratification, which one might consider the more logical definition of a Nusselt number. The latter includes, however,  a flux which is irrelevant for the convective flow  (the radiative heat flux due to the adiabatic part of the stratification);  it would contribute even if there is no flow at all\footnote{The significance of this distinction is not apparent in some of the numerical work on semiconvection, e.g.\ Biello (2001).}. \tbf{From (\ref{rgrad}), (\ref{rrha}), (\ref{fnu})} we obtain a relation between Nusselt number and density ratio for our astrophysical, imposed heat flux conditions:
\beq {\nabla_\mr{r}-\nabla_\mr{a}\over  \nabla_\mu}R_\rho=\nut.\eeq
To complete the model, a prescription for the Nusselt number as a function of the control parameters is needed. For this, we assume that the layer thickness $d$ is small enough that an incompressible approximation is valid, so the results of Sect.\ \ref{effd} can be used.  Eqs.\ (\ref{nusnut})--(\ref{Q}) then determine $\nut$ as a function of $\mr{Ra_*}$ and $R_\rho$. Together with  $R_\rho$ from (\ref{rrha}) and $\mr{Ra_*}$ from (\ref{raa}), this forms a set of equations for the superadiabaticity $\nabla-\nabla_\mr{a}$ as a function of the control parameters $d/H$ and $\nabla_\mr{r}$. 

\subsection{Asymptotic results}
This is a somewhat implicit algebraic problem, but the limiting case equivalent to $d/d_0\gg 1$ in Sect.\ \ref{fixed} again has  a simple form. This limit corresponds to 
\beq f^2(d/H)^4={g d^4\over \kt^2H}\gg 1,\eeq
or 
\beq d \gg l_0\equiv(\kt^2H/g)^{1/4}, \label{l0}\eeq
i.e. $d$ large compared to the length $l_0$ on which the thermal diffusion time scale equals the free fall time over a scale height.
As before, $R_\rho$ tends to its maximum, $R_\rho\approx\mr{Le}^{-1/2}$ in this limit, so that 
\beq \nut\approx {\nabla_\mr{r}-\nabla_\mr{a}\over\nabla_\mu}\,  \mr{Le}^{-1/2},\eeq
while the superadiabaticity follows from (\ref{rrha}):
\beq \nabla-\nabla_\mr{a}\approx\mr{Le}^{1/2}\nabla_\mu.\label{sa}\eeq
The solute Nusselt number follows as in Sect.\ \ref{fixed},
\beq \nus\approx \nut,\eeq
so the effective solute diffusivity is
\beq \kse=(\ks\kt)^{1/2}(\nabla_\mr{r}-\nabla_\mr{a})/\nabla_\mu,\label{nusef}\eeq
the same as in the simpler model of S92 and ZS10 \tbf{(apart from a factor involving radiation pressure)}. It is independent of the layering thickness $d$, within the range of validity of the approximations,
\beq l_0\ll d\ll H.\eeq
The relative thickness of the stagnant zones is
\beq \delta\approx 1/\nus=\mr{Le}^{1/2}\nabla_\mu/(\nabla_\mr{r}-\nabla_\mr{a}).\label{del}\eeq
For  Le $\ll \nabla_\mr{r}-\nabla_\mr{a}$ the stagnant zone is thus thin compared with the layer thickness  (but still significantly thicker than the boundary layers of the overturning zone, cf. Sect. \ref{stag}).

\section{Evolution of the layer thickness}
\label{evol}
The main uncertainty of any model for semiconvection is the thickness $d$ of the double diffusive layers. In a stellar interior the layer thickness has little influence on the quantities of interest (eqs. \ref{sa}, \ref{nusef}), but it can become important for the structure of the star when layer thickness becomes macroscopic, i.e. comparable with the pressure scale height. 

\begin{figure}
\center\includegraphics[width=0.6\hsize]{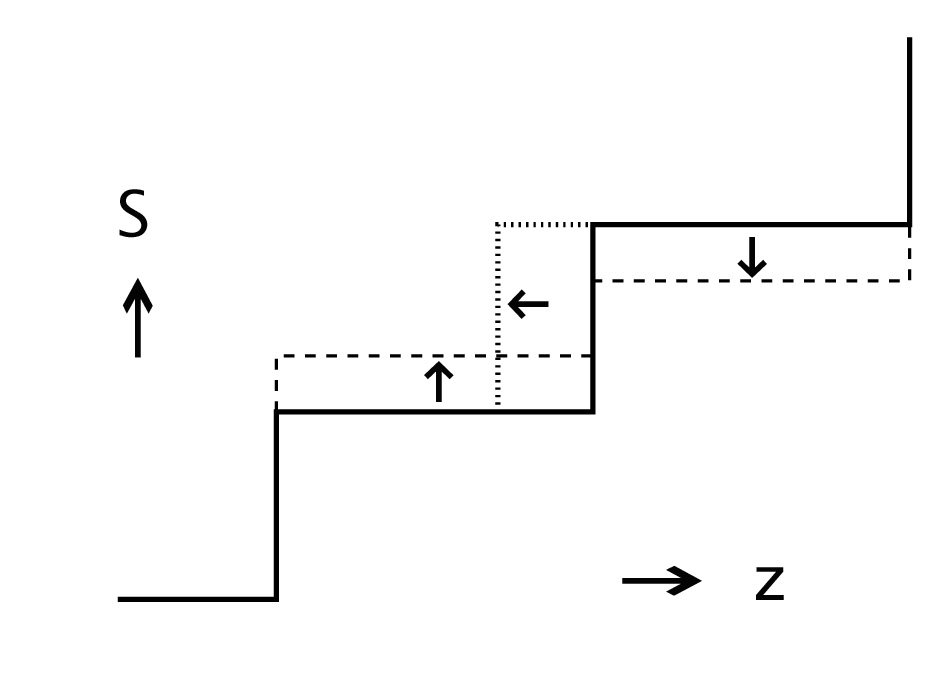}\\
\caption{Merging of two neighboring layers in a double diffusive \tbf{staircase}. The solute concentration can change by exchange between the two until the contrast disappears (dashed), or one can grow in thickness at the expense of the other (dotted).\label{merg}}
\end{figure}

In geophysical and laboratory cases, it is observed that $d$ evolves secularly, on a long time scale compared with the thermal diffusion time. Layer thickness is therefore a quantity that {\em cannot be discussed independently of the history of the system}.  \tbf{The process has not been studied very extensively (but see Wirtz \& Reddy 1979, McDougall 1981, Young \& Rosner 2000, Ross \& Lavery 2009, and the coffee table experiment in ZS13). }The layer thickness increases by a process of merging of neighboring layers. Two mechanisms are observed: vanishing contrast, and drift of interface position. This is illustrated schematically in Fig.\ \ref{merg}. 

Which of these two dominates, why, and at what rate the merging takes place appears to be not well understood. From the model presented here, however,  an estimate for the merging rate which is independent of this uncertainty  can be derived, as well as the resulting layer thickness as a function of time.

Both merging mechanisms involve the redistribution of solute between neighboring layers.  For example, if an interface between two layers transmits solute somewhat faster than average, the contrast between them decreases. The time scale $\tau$ on which such changes take place is limited by the effective solute diffusivity, hence is of the order
\beq \tau\approx d^2/\kse. \eeq
The rate of merging is then
\beq \mr{d}\ln d/\mr{d} t\approx1/\tau\approx\kse/d^2, \eeq
so that
\beq d\approx(2\kse t)^{1/2}\label{dmerg},\eeq
if  $\kse$ is constant in time.
The layer thickness is thus predicted to increase as the square root of time. \tbf{This agrees qualitatively with the dependence inferred empirically in some laboratory experiments. In most of these, however, the layering is induced by lateral heating  of the solute gradient rather than vertical heating, hence is not directly comparable with the astrophysical case. Wirtz \& Reddy (1979) for example find an initial square-root-like dependence, which saturates when the thickness reaches the separation between the lateral boundaries.} 

In terms of a merging time scale $t_\mr{m}$, (\ref{dmerg}) becomes
\beq t_\mr{m}\approx d^2/2\kse.\eeq
For a quantitative check, compare this to the measurements from lake Kivu, where the observed layer thicknesses are 0.3-0.5 m, the relative interface thickness $\delta$ of order 0.4  (Schmid et al. 2010). With the diffusivity of CO$_2$ in water, $\ks=2\,10^{-9}$ m$^2/$s, the effective solute diffusivity is 
\beq \kse=\nus\ks\approx\ks/\delta\approx7\, 10^{-9}~\mr{m}^2/\mr{s} .\eeq
For a layer of thickness 0.4 m this predicts a merging time of order 8 months. This agrees with the observed time scale for changes in the layering, of the order of several months. This comparison has to be taken with a grain of salt, of course, since the present analysis is not strictly valid for the Prandtl number of water.

\section{Summary}
The theory presented expands on the previous analyses in S92 and ZS10. It improves on these by including the effect of a {\em stagnant zone} of finite thickness. That is, the region over which heat and solute are transported by diffusion is not limited to just the boundary layers of the overturning interior of a double-diffusive step, but can in principle be an arbitrary fraction of the layer thickness. The need for this extension arose from observations of geophysical examples of thermohaline layering and results from numerical experiments.  

A simple 2-zone model consisting of a stagnant and an overturning zone, and using an experimental fitting formula for convective heat transport in the overturning zone produces a clear physical picture for the dependence of the effective transport  properties of double diffusive layered convection (Sect.\ \ref{effd}). The results predict the existence of a maximum  to the density ratio for which such a layered state can exist.  It is of the order $R_{\rho\,\mr{max}}\approx\mr{Le}^{-1/2}=(\kt/\ks)^{1/2}$ and approaches this value from below with increasing Rayleigh number (or layer thickness). 

Of special interest is the astrophysical case where the heat flux rather than a temperature difference is given. In this case the dependence of effective solute diffusivity on solute stratification and heat flux has the simple form (\ref{nusef}). This is the same as before in ZS10 and S92 (except for the effect of radiation pressure on the equation of state not included here). Due to the presence of the stagnant zone  the value of the superadiabaticity (\ref{sa}) differs from that in S92, ZS10. For practical conditions in a stellar interior, however, $\nabla-\nabla_\mr{a}$ is so small that its exact functional form makes little difference for the temperature stratification. The thickness of the stagnant zone is small relative to the layer thickness, as a consequence of the low value of the Lewis number. 

The main conceptual difference with respect to ZS10 and S92 is the existence of a maximum to the density ratio.  In the astrophysical case of imposed heat flux it plays only a rather implicit role, however. 
The differences are more significant when conditions are not in the astrophysically relevant limiting case. In numerical simulations in particular,  which are necessarily much closer to marginal conditions for double diffusive layering, the stagnant zone and the maximum density ratio have a substantial effect on the results. A comparison with such simulations is presented in the companion paper ZS13.

\tbf{Semiconvection is only one of the potential mixing processes in stars. Rotation induced mixing and magnetic processes are likely to be relevant as well, and could actually be more effective.}

\begin{acknowledgements} 
The author thanks F.\ Zaussinger for extensive discussions and suggestions, \tbf{and the referee for many corrections and detailed comments  which have led to significant improvement.}

\end{acknowledgements}

\end{document}